\documentstyle[preprint,tighten,aps]{revtex}
\begin{document}
\preprint{\vbox{\noindent
To appear in Phys. Rev. E 55 (1997)
\hfill nucl-ex/9612007\\
          \null\hfill  INFNCA-TH9613}}
\title{Search for non-Poissonian behavior in nuclear 
       $\bbox{\beta}$ decay
       }
\author{
         Giorgio Concas$^{1,2,}$\cite{email1}
     and Marcello Lissia$^{3,1,}$\cite{email2}
       }
\address{
$^{1}$Dipartimento di Scienze Fisiche, Universit\`a di Cagliari,
      via Ospedale 72, I-09124 Cagliari, Italy\\
$^{2}$Istituto Nazionale per la Fisica della Materia,
      via Ospedale 72, I-09124 Cagliari, Italy\\
$^{3}$Istituto Nazionale di Fisica Nucleare, Sezione di Cagliari,
      via Negri 18, I-09127 Cagliari, Italy\\
        }
\date{November 1996}
\draft
\maketitle 
\begin{abstract}
We performed two independent counting experiments on a $\beta$-emitting
source of $^{151}_{62}\text{Sm}$ by measuring the $\gamma$ photon emitted in
a fraction of the decays. For counting times ranging from $10^{-3}$ to 
$5.12\times 10^{4}$ seconds, our measurements show no evidence of deviations 
from Poissonian behavior and, in particular, no sign of $1/f$ noise. 
These measurements put strong limits on non-Poissonian components of the 
fluctuations for the subset of decays accompanied by $\gamma$, and 
corresponding limits for the total number of $\beta$ decays. In particular, 
the magnitude of a hypothetical flicker floor is strongly 
bounded also for the $\beta$ decay. This result further constrains theories 
predicting anomalous fluctuations in nuclear decays.
\end{abstract}

\narrowtext

\section{Introduction}
\label{sec:intr}

The statistics of the radioactive decay of heavy nuclei have been 
the subject of much experimental and theoretical work in the past 
decade. So wide an interest was stimulated by the conjecture that, 
owing to the intrinsic fluctuations of the decay rate, the counting 
statistics could depart from the simple Poissonian 
behavior~\cite{H75a,H75b,H80,VanVliet81,VanVliet82,VanVliet91,Hooge95}.

The experimental results are often conflicting, even for the same kind of 
source.  On the one hand, there exist investigations both on $\alpha$
($^{241}_{95}\text{Am}$~\cite{Prestwich86,Prestwich88,Pepper89,Kennett89} and
 $^{210}_{84}\text{Po}$~\cite{A89a}) and $\beta$ decays
($^{137}_{55}\text{Cs}$~\cite{Gopala88}) that confirm the Poissonian 
nature of these processes. 
 On the other hand, several experiments carried out both with $\alpha$ 
($^{241}_{95}\text{Am}$, $^{239}_{94}\text{Pu}$, and 
 $^{244}_{96}\text{Cm}$~\cite{Gong83,K86,K87,Rusov92}) and with
 $\beta$ sources ($^{204}_{81}\text{Tl}$~\cite{A89b},
 $^{90}_{39}\text{Y}$~\cite{A91} and 
 $^{90}_{38}$Sr-$^{90}_{39}$Y~\cite{Gopala94}), 
 find that the counting variance, for long counting periods, is higher than 
the Poissonian value by more than one order of magnitude.

This anomalous large variance has been taken as an experimental 
evidence that the power spectrum of the decay-rate fluctuations has 
a contribution that grows as the inverse of the frequency $f$ at 
low frequencies, in addition to the usual frequency-independent
Poissonian component.
Several mechanisms have been proposed as possible sources of 
this $1/f$ noise: quantum self-interference between the 
wave packets of the emitted particles~\cite{H80,VanVliet81}, 
solid-angle fluctuations and random rearrangements within the 
source~\cite{VanVliet91}, spatial $1/f$ noise in the 
detector~\cite{Rusov92}. As a matter of fact, the interpretation of 
the decay experiments reporting a variance in excess of the Poisson value 
is still an open problem~\cite{VanVliet91}.

In previous work~\cite{Boscaino94a,Boscaino94b} we considered the 
decay statistics of a $\gamma$ source ($^{119m}_{50}\text{Sn}$). In that case,
we measured that, for counting periods $T$ longer than one hour, 
the variance of the decay rate significantly deviated from the 
Poissonian prediction. However, that behaviour could be fully explained by 
taking into account the time dependence of the statistics~\cite{Teich79} 
without resorting to any exotic effect~\cite{Boscaino94a,Boscaino94b}.

The aim of this article is to extend our experimental study to a
different nucleus, $^{151}_{62}\text{Sm}$, that undergoes $\beta$ decay. 
There are in fact theoretical claims~\cite{VanVliet91} that deviations 
from Poissonian statistics could be caused by self-interference of the 
emitted particles and that these deviations should be present only in 
$\beta$ decays and not in $\gamma$ or $\alpha$ decays.

\section{Experimental setup}
\label{sec:E}
 
The mean lifetime of $^{151}_{62}\text{Sm}$ is ($130\pm 12$) 
years~\cite{ToI96}. While 
most of the nuclei directly $\beta$ decay into the ground state of 
$^{151}_{63}\text{Eu}$, a small fraction (0.91\%) $\beta$-decays into an 
excited state of energy $E_{\text{exc}} = (21.532\pm 0.068)$~keV, which then 
decays to the ground state (mean lifetime: $1.38\times 10^{-8}$ seconds). 
In the 3.45~\% of cases, this second fast transition produces a
$21.532$~keV $\gamma$-photon: our apparatus has been set up to detect this
photon. In summary, our measurement has the characteristic of selecting
a fraction $\xi$ of the total decays ($\xi = (3.14 \pm 0.22)\times 10^{-4}$):
those decays that go through the two-step process, $\beta$ emission followed 
by a $21.532$~keV photon~\cite{ToI96}. We shall discuss later and in the
Appendix why and to what extent our results on the statistics of the 
$\gamma$'s also carry information on the total statistics of the $\beta$ 
decays.

The source is a crystal of $\text{SmF}_3$ containing 
$^{151}_{62}\text{Sm}$ nuclei (activity: 3.7 GBq) shaped as a thin disk 
(diameter: 14 mm) with an aluminium cap. The aluminium cap, which closes 
the source, filters out the $\beta$ particles.
In the two experiments, which we denote A and B, we used the same source 
at different distances (about 15 and 7~cm, respectively) from the detector 
in order to change the count rate: while Poissonian statistics only depends
on the total number of counts (rate $\times$ time), deviations from the
standard case (and/or systematic errors) could in principle depend also 
on the rate (see Ref.~\cite{Boscaino94a} for one such an example) and it
is better to have the possibility of performing this kind of checks.

In each experiment the photons were detected by a disk-shaped crystal of 
NaI(Tl) (diameter: 5 cm) integrally mounted on a photomultiplier tube (PMT): 
we used a crystal 1 mm thick in experiment A, and a crystal 2 mm thick in 
experiment B.

In both experiments, the output signal from the PMT, after being amplified 
and shaped to a Gaussian pulse, passed through a single-channel 
analyzer, which selected pulses corresponding to an energy window
from 2 to 53~keV. The pulse-shaping time constant was
$0.5\mu$s and the time resolution of the single-channel analyzer 
$0.6\mu$s. The dead time of the entire system was about $2.5\mu$s.
The energy window was preliminarily set by means of a 
multi-channel analyzer module; we have verified that no appreciable drift 
of the window occurred during the experiments, which lasted 76 days (A) and
19 days (B). We verified that the stability of the energy window and of 
the voltage of the power supply were sufficient to keep systematic variations
of the counting rate below 0.01\%, therefore below the statistical
fluctuations we measured: only for the longest measurements
(total counts of the order of $10^8$) the fluctuations-to-signal ratio
was as low as $10^{-4}$ ($1/\sqrt{10^8}$).

Counting was executed by a programmable multi-channel scaler 
(MCS) module interfaced to an IBM PC, which provides for control 
and data storage. 
In experiment A (B), a set of 40 (38) values of the counting period 
$T$ was preliminarily defined in the control program with $T$ ranging
from $T_{\text{min}} = T_1= 10^{-3}$s to 
$T_{\text{max}} = T_{40}= 2^9\times 100$s 
($T_{\text{max}} = T_{38}= 2^7\times 100$s).
For each value of $T$ the MCS module counted the events occurring in each
of 64 consecutive periods of length $T$. Count data were saved on hard 
disk for further off-line analysis. At the end of experiment A (B), data 
were available as 40 (38) sequences of 64 counts $M^T_k$ ($k=1, \ldots, 64$),
one for each of the prefixed values of $T$.

\section{Results and  discussion}
\label{sec:R}
We analyzed the data by computing the average count and the Allan 
variance (see Refs.~\cite{VanVliet82,Prestwich91} and references therein) as 
function of the time interval $T$. All our results 
originate from a single uninterrupted run (for each experiment), have been 
averaged over the same number (64) of consecutive intervals, and are 
statistically independent (each count has been used only once).

First we verified that the count rate during each experiment had no drifts
that could bias the Allan variance; in particular, the slow exponential decay
of the source could not affect the Allan variance at the low count rates 
we operated~\cite{Boscaino94a}. Therefore, it is consistent that
we consider a constant average rate. We measured this average rate
\begin{equation}
 m = \sum_{T} \frac{1}{T} \frac{1}{64} \sum_{k=1}^{64} M^T_k \quad ,
\label{def:rate}
\end{equation}
finding $m = (5.3687 \pm 0.00025) \times 10^3$~count/s in experiment A 
and $m = (2.4262 \pm 0.00022) \times 10^4$~count/s in experiment B.
Being the rate constant, the average count for an interval of length $T$
\begin{equation}
   \overline{M}(T) \equiv \frac{1}{64} \sum_{k=1}^{64} M^T_k
\label{def:ave}
\end{equation}
has an expectation value proportional to $T$: $\langle \overline{M}(T) 
\rangle= m T$.

There are two reasons for using the Allan variance, which we estimate with
an average over 63 consecutive measurements
\begin{equation}
   A(T) \equiv 
\frac{1}{2\times 63}\sum_{k=1}^{63} [M^T_k - M^T_{k+1} ]^2 \, ,
\label{def:All}
\end{equation}
instead of the usual variance.
The first and most important reason is that $A(T)$ is finite even when the
power spectrum grows as $1/f$ at low frequencies: when non-Poissonian 
fluctuations might be present, the Allan variance is then a common choice.
We remind that the power spectrum of Poissonian fluctuations
is independent of frequency: $S(f)= 2 m T$; since counts are uncorrelated,
the Allan and the usual variance have the same expectation (use
$\langle M^T_j M^T_k\rangle \propto \delta_{j k}$ and Eq.~(\ref{def:All})), 
namely, the average count: 
$\langle A(T) \rangle = \langle\overline{M}(T) \rangle = m T$.
However, if the fluctuations have a power spectrum $S(f) = C/ f$
($C$ is a constant independent of $f$), i.e., we are in presence of
$1/f$ noise, the expectation value of $A$ is: 
$\langle A(T) \rangle = C \ln 4\, T^2 $
(note the different power of $T$ compared to Poissonian fluctuations), 
while the usual variance is infinite~\cite{VanVliet82}. A second additional 
advantage of using the Allan variance is that it is less sensible to drifts of
the count rate: the correction is independent of the number of intervals
(64) and not proportional to it, see the Appendix of Ref.~\cite{Boscaino94a}.

Before discussing our results, we wish to comment on our choice of
observing the channel of the decay characterized by the emission
of a 21.532~keV photon. A more detailed discussion can be found in the
Appendix. We made this choice because we can control
better the stability of our measurements when detecting photons than
when detecting electrons, given our present equipment. However, since
one of the motivations of our experiment was to study fluctuations in a
$\beta$ decay, it is natural to ask to what extent the statistics of the 
$\gamma$ emission reflects the statistics of the $\beta$ decay.
The time delay of the emission is so small (mean lifetime of the
excited state: $1.38\times 10^{-8}$ seconds) compared to the time intervals
of interest that its effect is negligible. Yet one might worry 
that the fluctuations of the small branching ratio (the fraction of decays
that on average emit the $\gamma$ is only $\xi =0.000314$) might overwhelm any
exotic effect of the original decay. The explicit calculation reported in
the Appendix shows that: (1) an upper bound on the flicker floor in the
statistics of the $\gamma$'s implies an equal bound on the flicker floor
in the statistics of the parent decay; (2) an upper bound on the
ratio of $1/f$ noise to Poissonian noise in the statistics of the 
$\gamma$'s implies a corresponding bound for the statistics of the parent 
decay weaker by a factor $1/\xi$; (3) upper bounds on less singular, e.g. 
frequency independent, deviations from Poissonian behavior in the 
statistics of the $\gamma$'s imply corresponding bounds on the parent 
decay: these bounds on the statistics of the $\beta$ decay are also weaker 
by a factor $1/\xi \approx 3000$.

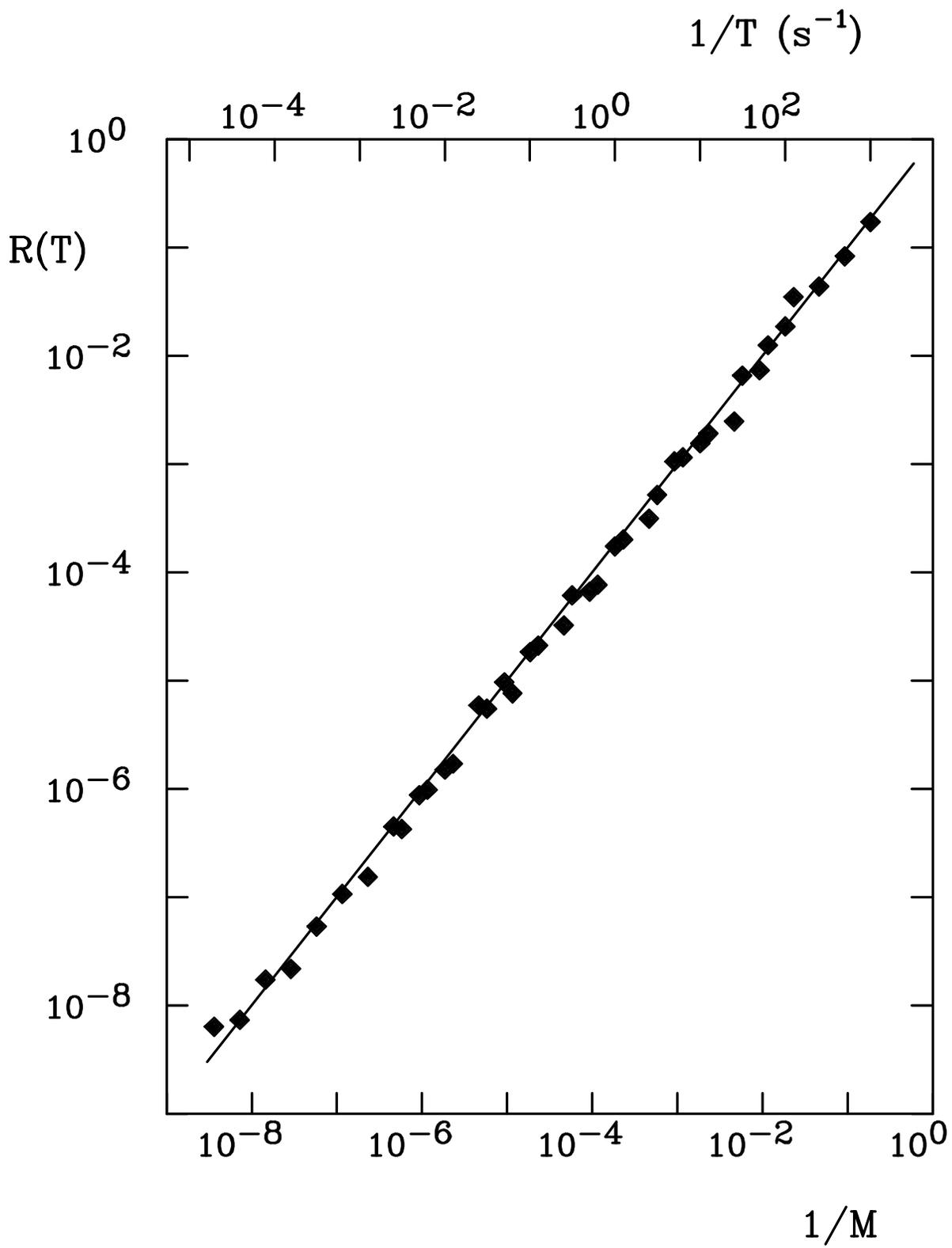
\begin{figure}
\caption[aaa]{Relative Allan variance $R(T)$ versus the inverse mean number
              of decays $1/\overline{M}$ (lower scale) and the inverse 
              time interval $1/T$ (upper scale) for the experiment A 
              (average rate $m\equiv M/T = 5.3687\times 10^3$). 
              Diamonds are the experimental values. The solid line is the 
              Poissonian prediction $R = 1/M$. }
\label{fig1}
\end{figure}

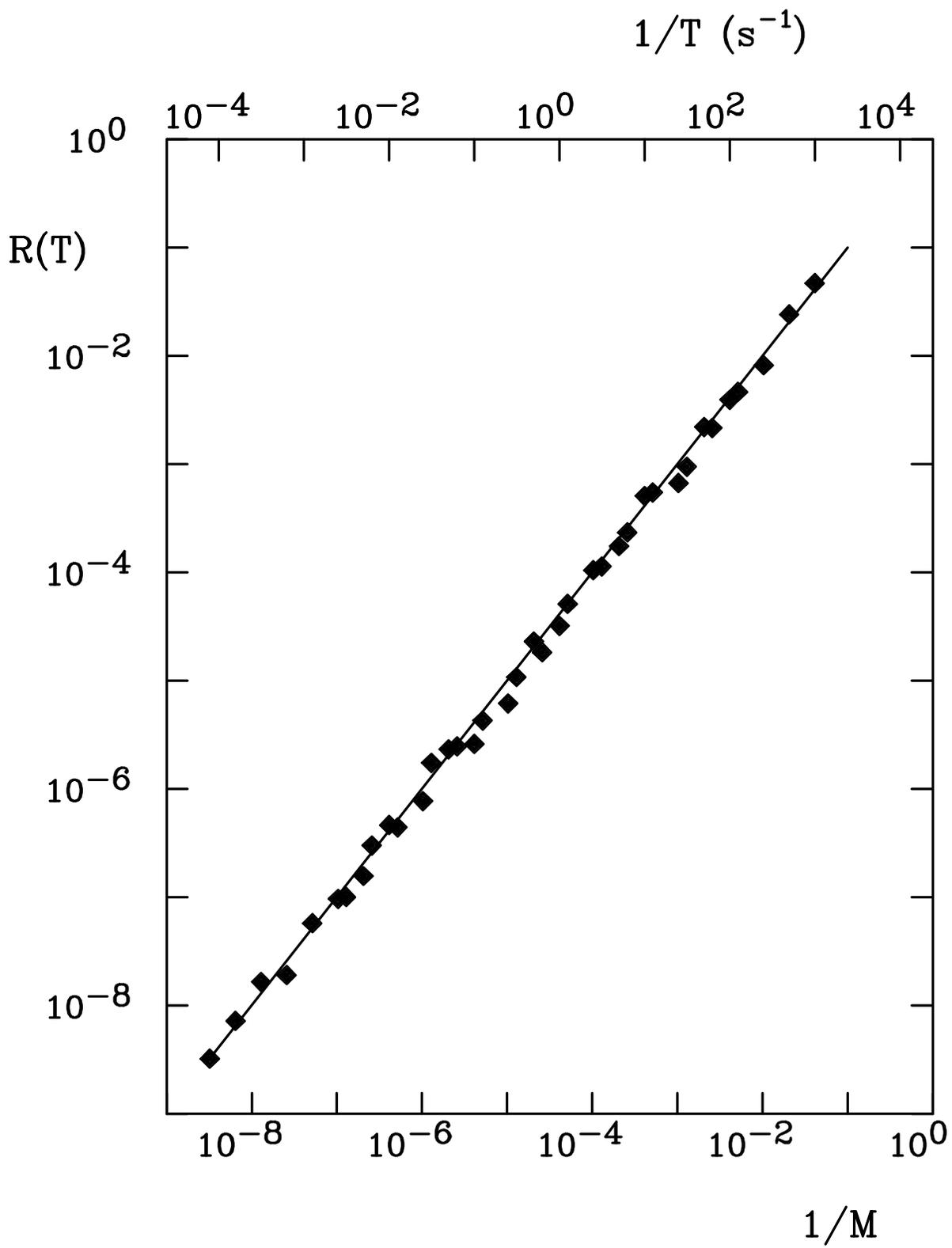
\begin{figure}
\caption[bbb]{Same as Fig.~\ref{fig1} for the experiment B
              (average rate $m = 2.4262\cdot 10^4$).}
\label{fig2}
\end{figure}

We report in Figs.~\ref{fig1} and \ref{fig2} the ratio 
$R(T)\equiv A(T)/M^2(T)$ (reduced Allan variance) versus the inverse of
the number of counts $1/M(T)$ for experiments A and B, respectively.  
Both experiments show that $R(T)$ depends linearly on $1/M(T) = 1/(m T)$ 
with unit slope in the range of $T$ considered.

The data perfectly fit the Poisson prediction $R(T)=M/M^2 =1/M\propto 1/T$;
this prediction is also reported in Figs.~\ref{fig1} and \ref{fig2} 
as a solid line. A fit to the data yields $M(T)\times R(T) = 0.99\pm 0.02$.

On the contrary, a power spectrum $S(f) = C / f$ would yield
$R(T)= (C \ln 4\, T^2) / (m T)^2 = C \ln 4 / m^2 $. Therefore, if we suppose 
that both a Poissonian and a $1/f$ contribution are present, when $T$ is large 
enough the Poissonian contribution becomes negligible and $R(T)$ goes
to a constant ($F\equiv C \ln 4 / m^2$): this constant $F$ is usually called 
flicker floor. We measured values of $R(T)$ as low as $6 \times10^{-9}$ 
($3 \times10^{-9}$) in the experiment A (B) at 
$T_{\text{max}} = 5.12 \times 10^4$~s ($T_{\text{max}} = 1.28 \times 10^4$~s) 
without seeing deviations from Poissonian behavior and, in particular, no 
signal of the curve turning up at large $T$ and becoming constant. Therefore,
we conclude that, if a flicker floor is present,
$F<3 \times10^{-9}$; as discussed in the Appendix, this limit is valid
also for the $\beta$ decay.

If we express the power spectrum as the sum of the Poissonian component 
plus a hypothetical $1/f$ component: $S(f) = 2 m T + C/f $, in the range of 
frequencies accessible by our experiments ($f>1/T_{\text{max}}$), the
limit on the flicker floor implies an upper limit on the ratio of the strength 
of the $1/f$ contribution ($C/f$) relative to the Poissonian one ($2 m T$), 
i.e., a limit on the ratio $(C/f) / (2 m T_{\text{max}}) < C/(2 m) $:
$(C/f)/(m T_{\text{max}}) < 1 \times 10^{-5}$ 
($C/m < 2.5\times 10^{-5}$). These limits on the strength of the $1/f$ noise
are valid for the channel of the decay with $\gamma$ emission; for the total 
$\beta$ decays the limit is weaker (see Appendix): 
$(C_{\beta}/f)/(m_{\beta} T_{\text{max}}) < 3 \times 10^{-2}$.

The model of quantum $1/f$ noise proposed by Handel predicts
$F = 8\alpha \zeta\ln2\,  (\Delta v/c)^2/(3\pi)$ for $\beta$ decays,
see Eq.~(3.6) of Ref.~\cite{VanVliet82}, and references therein; 
here $\alpha\approx 1/137$ is the fine structure constant, $0<\zeta<1$ 
is a coherence factor, and $\Delta v/c$ is the velocity change of the 
particles in the emission process relative to the speed of light $c$:
if $K_{\beta}$ is the kinetic energy of the electron, 
$(\Delta v/c)^2 = 1-[1+K_{\beta}/(m c^2)]^{-2}$. Since we did not
measured the electron energy, our data are averaged over the entire
electron-energy spectrum. Therefore, we can only give an estimate of
the limit on the coherence factor by using the average electron
energy: $\langle K_{\beta}\rangle =13.96$~keV. The fact that we do not see 
any flicker floor implies, in the context of Handel's model, that the 
coherence factor $\zeta$ must be smaller than about $10^{-5}$. Our limit 
should be compared to the recent positive determinations of $\zeta$ in
the range $5.2 \times 10^{-3}< \zeta < 8.3 \times 10^{-3}$, that Gopala 
et al.~\cite{Gopala94} have made, albeit in different $\beta$ decays:
$^{90}_{38}$Sr, $^{90}_{39}$Y and $^{204}_{81}$Tl. We do not have any 
explanation why $\zeta$ should be more than two orders of magnitude 
larger in those decays compared to our upper limit.

\section{Conclusions}
\label{sec:C}

     We have measured the counting rate of secondary $\gamma$ rays from
a $\beta$ source of $^{151}_{62}\text{Sm}$ for counting periods ranging 
from $10^{-3}$ to $5.12\times 10^4$ seconds, and studied the fluctuations 
of the rate by means of the Allan variance.
\begin{itemize}
\item
We have found no evidence of deviations from Poissonian behavior up to
a ratio of fluctuations to signal as low as $ 5 \times 10^{-5}$.

\item
The ratio between a hypothetical $1/f$ component of the power spectrum 
and the usual Poissonian contribution must be less than $1\times 10^{-5}$ 
at the longest time interval (lowest frequency) that we have measured 
($T_{\text{max}} = 5.12\times 10^4$~s).

\item
We found no evidence of flicker floor. The upper bound on a hypothetical 
flicker floor is $ 3 \times 10^{-9}$; this limit is valid also for the
statistics of the total $\beta$ decays.

\item
If our upper bound the flicker floor is interpreted in the context of 
Handel's theory of $1/f$ noise predicting coherent interference of the 
emitted charged particle, the coherence factor $\zeta$ for this decay must 
be less than about $1\times 10^{-5}$: this number is more than two orders 
of magnitude smaller than the one that has been recently proposed in the 
literature albeit for different decays~\cite{Gopala94}.
\end{itemize}

\acknowledgments
We gratefully acknowledge stimulating and encouraging discussions with 
R.~Boscaino.

This work has been partially supported by M.U.R.S.T. (Italian Ministry
of University and Scientific and Technological Research).

\appendix
\section*{}
In this Appendix, we discuss the relation between fluctuations of the number 
of total decays (in our case, the total number of $\beta$ decays) and
fluctuations of the number of decays in a subchannel (in our case, the 
fraction of $\beta$ decays that produce a $\gamma$ photon with 
energy 21.532~keV).

The main results of this Appendix are summarized by Eqs.~(\ref{AgAb}) 
    and (\ref{RgRb}).

We consider only the effect of the fluctuations of the branching ratio and
not the effect of time delay between the first and second decay, which 
in general has the effect of a low-pass filter~\cite{Prestwich88}, since this 
second decay (the $\gamma$ emission) is practically instantaneous for the case 
under study (mean lifetime 13.8~ns compared to a time resolution of the 
order of $\mu$s and to the shortest time interval considered: 1~ms).

In the following we shall use the symbol $M$ when referring to the number of
detected $\gamma$'s and the symbol $N$ when referring to the corresponding
(total) number of $\beta$ decays. We shall also use the subscript $\gamma$ 
($\beta$) referring to the partial daughter statistics (total parent 
statistics). Let us define two kinds of averages:\\
$\langle \cdots \rangle_{N}$ average over the $\gamma$-count distribution 
   keeping the number of $\beta$ counts $N$ fixed;\\
$\bbox{\langle} \cdots \bbox{\rangle}$ average over the $\beta$-count
  distribution.\\
Using the symbol $\xi$ for the branching ratio, we can write
\begin{eqnarray}
\langle M \rangle_{N} &=& \xi N \\
\overline{M}\equiv \bbox{\langle}  \langle M \rangle_{N} \bbox{\rangle} 
   &=& \xi \overline{N} \quad ,
\end{eqnarray}
where for simplicity we use the symbol $\overline{M}$ to indicate the number
of $\gamma$ counts twice averaged both over the $\gamma$ and $\beta$ 
distributions and, at the same time, the symbol $\overline{N}$ to indicate
the average $\beta$ counts (over the $\beta$ distribution).

Since our experiments do not show any deviation from Poissonian behavior,
we can readily put limits on non-Poissonian components of the $\gamma$
counts. The implications for the total $\beta$ decay can be assessed by
assuming that the $\gamma$ distribution at fixed number of $\beta$ decays
$N$ is standard (binomial and frequency independent) and by considering the 
effect of fluctuations of $N$:
\begin{equation}
\label{MiMj}
\langle ( M_{i} - \langle M \rangle_{ N_{i} } ) 
        ( M_{j} - \langle M \rangle_{ N_{j} } ) 
\rangle_{ N_{i} N_{j} }
   = \delta_{ij} \langle  M \rangle_{ N_{i} }
   = \delta_{ij} \xi (1-\xi) N_{i} \quad ,
\end{equation}
where the indices $i$ and $j$ indicate different counting intervals.

We first consider the average at fixed $N_{i}$ and $N_{i+1}$ of
$( M_{i} - M_{i+1} )^2$, which by adding and subtracting
$\langle M \rangle_{N_i} =\xi N_i $ can be written as
\begin{eqnarray}
\langle ( M_{i} - M_{i+1} )^2 \rangle_{ N_{i} N_{i+1} } 
&=& \langle\,
\bbox{[}
 ( M_{i} - \langle M \rangle_{ N_{i} } ) -
 ( M_{i+1} - \langle M \rangle_{ N_{i+1} } ) +
 \xi ( N_{i+1} - N_{i} )
\bbox{]}^2
    \,\rangle_{ N_{i} N_{i+1} } \\
&=& \Bigl\{
      \langle ( M_{i} - \langle M \rangle_{ N_{i} } )^2 \rangle_{ N_{i} }
    + \langle ( M_{i+1} - \langle M \rangle_{ N_{i+1} } )^2 \rangle_{ N_{i+1} }
    + \xi^2 ( N_{i} - N_{i+1} )^2  \nonumber\\
\label{Moffd}
&&  - 2 \langle 
          ( M_{i} - \langle M \rangle_{ N_{i} } )
          ( M_{i+1} - \langle M \rangle_{ N_{i+1} } )
        \rangle_{ N_{i} N_{i+1} } \\
&&  + 2 \xi ( N_{i} - N_{i+1} ) 
           \,\bigl[
       \langle ( M_{i} - \langle M \rangle_{ N_{i} } ) \rangle_{ N_{i} }
     + \langle ( M_{i+1} - \langle M \rangle_{ N_{i+1} } ) \rangle_{ N_{i+1} }
           \bigr]
    \Bigr\} \nonumber \\
&=& \xi(1-\xi) (N_i+N_{i+1}) + \xi^2 ( N_{i} - N_{i+1} )^2 \quad ,
\end{eqnarray}
where we have applied Eq.~(\ref{MiMj}) to the first and second line of 
Eq.~(\ref{Moffd}), while the third line is identically zero.

If we now divide the above result by two and average it over the $\beta$
distribution, we find (considering that for a stationary process
$\bbox{\langle}  N_i  \bbox{\rangle} = \overline{N}$ independently of $i$):
\begin{equation}
\label{Allangb}
\frac{1}{2} \bbox{\langle} \langle 
        ( M_{i} - M_{i+1} )^2
       \rangle_{ N_{i} N_{i+1} } \bbox{\rangle} =
 \xi(1-\xi) \overline{N} + \xi^2 \frac{1}{2} \bbox{\langle}
                           ( N_{i} - N_{i+1} )^2 
                           \bbox{\rangle}\quad .
\end{equation}
The left-hand side of Eq.~(\ref{Allangb}) is the expectation value of the
Allan variance of the $\gamma$ counts, which we measure with the statistics
defined in Eq.~(\ref{def:All}), while the right-hand side is the expectation
value of the Allan variance of the total $\beta$-decay counts. If we
define $A_{\gamma}$ ($A_{\beta}$) as the Allan variance and
$R_{\gamma}\equiv A_{\gamma}/\overline{M}^2$ 
($R_{\beta}\equiv A_{\beta}/\overline{N}^2$) as the relative Allan variance
of the $\gamma$ ($\beta$) counts, Eq.~(\ref{Allangb}) becomes
\begin{eqnarray}
\label{AgAb}
A_{\gamma} &=& (1-\xi) \overline{M} + \xi^2 A_{\beta} \\
\label{RgRb}
R_{\gamma} &=& \frac{(1-\xi)}{\overline{M}} + R_{\beta} \, ,
\end{eqnarray}
which constitute the main result of this Appendix. 

In the following we analyzes the consequences of Eq.~(\ref{RgRb}) for our 
experimental study.

\subsubsection{Flicker floor}
If the $\beta$ decay has a $1/f$ component that produces a flicker floor
$F_{\beta}$ in the relative Allan variance, i.e., 
$R_{\beta}=1/\overline{N} + F_{\beta}$, the fact that no deviation of
$R_{\gamma}$ from $1/\overline{M}$ has been observed for $R_{\gamma}$ as
low as $3\times 10^{-9}$ implies not only that $F_{\gamma}< 3\times 10^{-9}$,
but also that $F_{\beta}< 3\times 10^{-9}$, where we have used 
Eq.~(\ref{RgRb}) dropping $\xi\approx 3\times 10^{-4}$ compared to 1: 
$(1-\xi) \approx 1$ and $1/\overline{M} + 1/\overline{N}  
  \approx 1/\overline{M}$. 

\subsubsection{$1/f$ noise}
If we are interested in the ratio of the $1/f$ contribution
($C_{\beta} / f$) relative to the Poissonian one ($2 m_{\beta} T $), we
should recall that the rate of the $\beta$ decay is 
$m_{\beta} = m_{\gamma}/\xi$ and the that the constant $C$ is related to 
the flicker floor by $C = F\times m^2 /ln4$.
Then this ratio for the $\beta$ decay can be related to the same ratio
for the $\gamma$ decay by using the fact that the limit on $F_{\beta}$ and
the one on $F_{\gamma}$ are equal: 
$(C_{\beta}/f) /( 2m_{\beta} T_{\text{max}} ) =
 (F_{\beta} m_{\beta} /f) /( T_{\text{max}} \ln4) =
 (1/\xi)\times (F_{\gamma} m_{\gamma} /f) /( T_{\text{max}} \ln4)$.
We loose a factor $1/\xi \approx 3000$ going from the upper bound on the
ratio of the $1/f$ contribution relative to the Poissonian one for the
$\gamma$ statistics ($1\times 10^{-5}$) to the upper bound on the same ratio 
for the $\beta$ statistics ($3\times 10^{-2}$).

\subsubsection{Frequency-independent non-Poissonian component}
If instead we suppose that the $\beta$ decay has a frequency-independent
deviation from Poissonian statistics, i.e.,
$R_{\beta}=\kappa/\overline{N} = \kappa\xi /\overline{M}$, the fact that
no deviation of $R_{\gamma}= (1-\xi+\kappa\xi)/\overline{M}$ from 
$1/\overline{M}$ has been observed ($\overline{M} \times R_{\gamma} 
= 0.99\pm 0.02$) implies also that $(1-\kappa)\xi= 0.01 \pm 0.02$ 
and, consequently, that $\kappa = -30 \pm 60 $.

\medskip
In conclusion, we have show in this Appendix that a measurement of a
process ($\beta$ decay) by selecting a subprocess (detecting the $\gamma$
emitted in a fraction of the decays) whose branching ratio
$\xi$ is itself a statistical variable corresponds, as might have been 
expected, to the use of a detector with efficiency not greater than $\xi$.
Therefore, we loose a factor $1/\xi$ in most limits on dimensionless 
quantities when passing from statistics of the subprocess to total 
statistics of entire process. However, there exist quantities, such as
the flicker floor, that can be determined from the partial statistics
without loosing any sensibility. The reason of this different behavior is
related on how strongly the noise under study depends on the number of
events $N$ compared to the usual $\sqrt{N}$ dependence.

\newpage

\end{document}